# Reconstructing Undersampled Photoacoustic Microscopy Images using Deep Learning

Anthony DiSpirito III, Daiwei Li, Tri Vu, Maomao Chen, Dong Zhang,
Jianwen Luo, Roarke Horstmeyer, and Junjie Yao

*Abstract*—One primary technical challenge in photoacoustic microscopy (PAM) is the necessary compromise between spatial resolution and imaging speed. In this study, we propose a novel application of deep learning principles to reconstruct undersampled PAM images and transcend the trade-off between spatial resolution and imaging speed. We compared various convolutional neural network (CNN) architectures, and selected a fully dense U-net (FD U-net) model that produced the best results. To mimic various undersampling conditions in practice, we artificially downsampled fully-sampled PAM images of mouse brain vasculature at different ratios. This allowed us to not only definitively establish the ground truth, but also train and test our deep learning model at various imaging conditions. Our results and numerical analysis have collectively demonstrated the robust performance of our model to reconstruct PAM images with as few as 2% of the original pixels, which may effectively shorten the imaging time without substantially sacrificing the image quality.

*Index Terms*—convolutional neural networks, deep learning, Fully Dense U-net, murine brain vasculature, photoacoustic microscopy, undersampled images, high speed imaging

Manuscript received May 29, 2020; date of current version May 29, 2020. This work was supported in part by the National Institutes of Health (R01 EB028143, R01 NS111039, R01 NS115581, R21 EB027304, R43 CA243822, R43 CA239830, R44 HL138185); Duke MEDx Basic Science Grant; Duke Center for Genomic and Computational Biology Faculty Research Grant; Duke Institute of Brain Science Incubator Award; American Heart Association Collaborative Sciences Award (18CSA34080277).

A. DiSpirito III, D. Li, T. Vu, M. Chen, D. Zhang, and J. Yao are all with the Photoacoustic Imaging Lab, Duke University, Durham, NC 27708 USA (email: jy211@duke.edu).

D. Zhang and J. Luo are with the Department of Biomedical Engineering, Tsinghua University, Beijing, China, 100084.

R. Horstmeyer is with the Computational Optics Lab, Duke University, Durham, NC 27708 USA

## I. INTRODUCTION

PHOTOACOUSTIC microscopy (PAM) is a hybrid imaging modality that combines optical excitation and ultrasonic detection [1, 2, 3]. In PAM, a pulsed laser provides excitation light that is absorbed by biological tissues. The photothermal effect induces a temperature rise that generates a pressure rise via thermo-elastic expansion that is proportional to the original optical absorption [2]. This pressure rise propagates as ultrasound waves, which are detected by an ultrasonic transducer to form an image of the original optical energy deposition inside the tissue. PAM utilizes either tightly or weakly focused optical excitation and focused ultrasound detection, combined with point-by-point raster scanning, to form high-resolution three-dimensional (3D) images [4].

In PAM, there is usually a trade-off between the imaging speed and the spatial resolution. To ensure the high spatial resolution, the raster scanning step size in PAM needs to be no larger than half of the expected spatial resolution, according to the Nyquist sampling theorem, similar to other pure optical microscopy technologies. However, pure optical microscopy can use air-based optical scanners (*e.g.*, MEMS mirror and galvo scanner) and easily achieve a high imaging speed over a large field of view. PAM, conversely, requires simultaneous scanning of both the optical excitation beam and the resultant ultrasound waves in an aqueous environment [4]. This restriction results in the low imaging speed of traditional PAM systems that mostly use slow mechanical scanning methods, especially when a large field of view is imaged without sacrificing the spatial resolution. If undersampling is performed with a large scanning step size, the imaging speed can be improved, but at the cost of the spatial resolution and thus image quality.

There have been recent efforts toward improving PAM imaging speed via advanced scanning mechanisms, such as water-immersible MEMS and polygon scanners [2, 4]. In addition, compressive sensing methods like single pixel and digital micromirror devices have been explored in PAM as an avenue to speed up imaging [5, 6, 7, 8]. However, these methods often need modifications of the system hardware and are not available for a traditional PAM system.

There have been a number of deep learning applications in photoacoustic computed tomography (PACT) to remove artifacts [9, 10, 11, 12, 13, 14, 15, 16, 17] and improve contrast [18, 19, 20] from undersampled data. Unlike PACT, PAM uses direct image formation without inverse image reconstruction [21, 22, 23, 24]. This absence of inverse image reconstruction has led the application of deep learning in PAM to be scarce so far, with one example using the technique for PAM motion-correction [25]. Outside of deep learning, dictionary learning has recently been reported to reconstruct undersampled PAM images [26]. However, dictionary learning often learns far fewer parameters than deep neural networks and lacks the benefit of layered operations. Thus, there is still a strong need for novel methods that can improve the imaging speed of traditional PAM systems without deteriorating the image quality or increasing the system complexity.

In this paper, we propose a deep learning approach to improve undersampled PAM images, using as few as 2% of the original pixels. Our deep learning technique offers an improved ability to approximate the nonlinear mapping of the undersampled images to their fully-sampled counterparts. Our method differs from previous efforts in high-speed scanners because it offers a software-only solution to the resolution-speed tradeoff. Moreover, our deep learning model was trained on a large number of fully-sampled PAM images as the ground truth, and thus we were able to circumvent the obstacle of validating the *in vivo* testing results.

## II. METHODS

### A. Deep Learning Framework

We first assume in Eq. (1) that there exists an approximate function *F* that maps the undersampled PAM image, $\mathbf{X} \in \mathbb{R}^{m \times n}$, to the fully-sampled PAM image, $\mathbf{Y} \in \mathbb{R}^{u \times v}$. We then train our deep learning model to learn a mapping $G(\boldsymbol{\theta},\mathbf{X})$, with both a parameter matrix $\boldsymbol{\theta}$ and the downsampled image $\mathbf{X}$ as input, such that our specified loss function is minimized via supervised learning so $G(\boldsymbol{\theta},\mathbf{X}) \approx F(\mathbf{X}) = \mathbf{Y}$.

$$F: X \to Y \quad (1)$$

### B. Loss Function

For the primary loss function, representing the pixel-wise error, we use the mean absolute error (MAE) between the ground truth image $\mathbf{Y}_{true}$ and the reconstructed image $\mathbf{Y}_{recon}$:

$$L_{MAE} = \frac{1}{N}\sum_{i=1}^{N}|Y_{true} - Y_{recon}| \quad (2)$$

Similar to [27], we also use Fourier loss (FMAE) where the mean absolute error is calculated from the magnitude of the 2D Fourier transform of $\mathbf{Y}_{true}$ and $\mathbf{Y}_{recon}$:

$$L_{FMAE} = \frac{1}{N}\sum_{i=1}^{N}||\mathcal{F}(Y_{true})| - |\mathcal{F}(Y_{recon})|| \quad (3)$$

The MAE and FMAE are combined via a weighted sum:

$$L_{Total} = \lambda_1 L_{MAE} + \lambda_2 L_{FMAE} \quad (4)$$

We use the weights of $\lambda_1 = 1.0$ and $\lambda_2 = 0.01$, respectively. The pixel-wise loss in the Fourier domain provides the optimizer more information about the vessel orientations and may highlight remnants of uniform downsampling as frequency corruptions. However, because the FMAE loss can contribute to training instability (especially during the early iterations) [27], we have decided to use a small weighting factor to limit its overall impact on the loss function.

### C. Fully Dense U-net

Fully Dense U-net (FD U-net), first proposed by Guan *et al.* [11] and later used by Nguyen *et al.* [27] and Vu *et al.* [16], implements dense blocks in both the expanding and contracting paths of U-net. These dense blocks allow each layer to learn additional feature maps based on the knowledge gained by all previous layers. By doing this, FD U-net effectively allows each layer within the dense block to build on each other, as the input to each layer in a dense block is concatenated with the outputs of all previous layers [11]. This ensures that each layer only needs to learn refinements that either augment the previous layers or diversify the collective feature set [11]. In addition, these dense blocks also allow for deeper networks without the issue of a vanishing gradient [11].

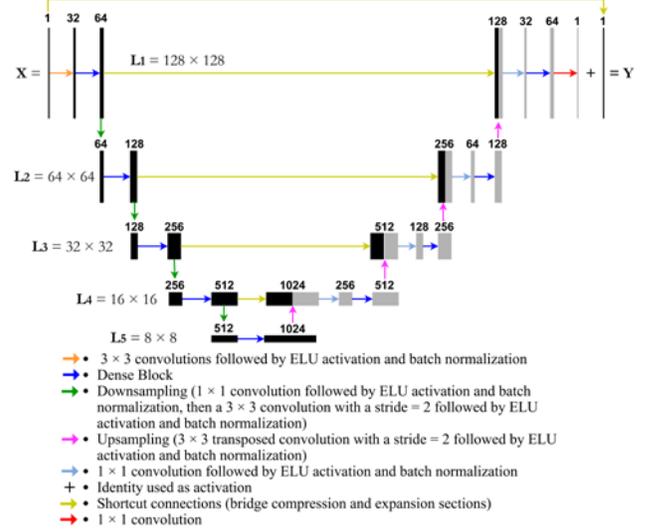

Fig. 1. Visual representation of FD U-Net architecture used as the basis for our model. The variables $L_1$, $L_2$, $L_3$…$L_i$ refer to the level of compression depth (i) within the model and the image size (N × N) at that compression depth.

As shown in Fig. 1, our implementation of FD U-net has two modifications from the original model: 1) RELU activation is replaced with ELU activation and 2) the max pooling layers are replaced with a 1 × 1 convolution block and a 3 × 3 convolution block with a stride of 2. The first modification benefits from batch normalization in mitigating an exploding gradient, and has been shown to improve learning speed in deeper residual networks [28, 29]. The second modification allows for a learned downsampling operation rather than the rigid max pooling procedure. Similar to the original FD U-net architecture, all convolutional blocks in our model have a batch normalization block following the convolution and activation. The fully dense blocks follow the same structure as Guan *et al.* [11], as illustrated in Fig. 2.

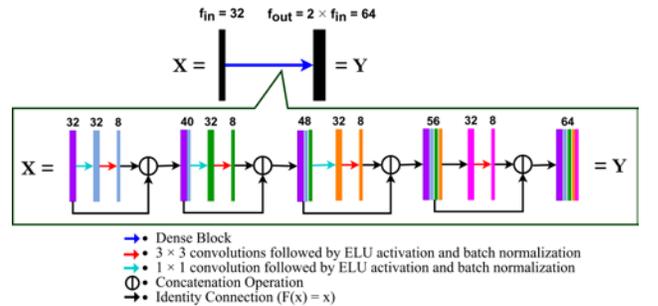

Fig. 2. A detailed view of an FD U-Net Dense Block. **X** is the input to the Dense Block and **Y** is the output. **f** is the number of filters. $f_{out}= 2 \times f_{in}$; for $f_{in} = 32$, $f_{out} = 64$. **k** = 8 as **k** = $f_{in}/4$.

## III. EXPERIMENTAL PROCEDURES

### A. Data Preparation

Our dataset is composed exclusively of *in vivo* mouse brain microvasculature data acquired by the Photoacoustic Imaging Lab at Duke University, using the PAM system previously published in [30]. This PAM system has a lateral resolution of 5 μm and an axial resolution of 15 μm. We are particularly interested with *in vivo* mouse brain imaging, because (1) PAM has been playing an increasingly important role in neuroscience, (2) functional brain imaging needs a high imaging speed, and (3) mouse brain vasculature has clear organization and patterns to be learned [31, 32, 33, 34, 35, 36, 37, 38] . Our dataset contained 292 maximum-amplitude-projection images of mouse brain vasculature, all acquired at a wavelength of 532 nm. We applied thresholding, a $3 \times 1$ median filter, and a $1 \times 3$ median filter to remove the noise and artifacts in the images. In addition, we performed contrast enhancement and grayscale intensity rescaling. The data was then randomly divided into approximately 80 percent training (233 images), 10 percent validation (30 images), and 10 percent testing (29 images). The training dataset was used in the optimization of the stochastic gradient descent algorithm, the validation dataset was used during training to save the best model according to the validation metrics, and the testing dataset was reserved to compare model inference performance.

### B. Data Augmentation

Our dataset is composed of images that are greater than 128-by-128 pixels, but our model expects an input image size of 128-by-128. First, all images were zero padded so that the pixel sizes of the images were evenly divisible by 128 in both the x and y directions. Our model during the inference phase worked with 128-by-128-pixel patches from images of various sizes that are similarly padded (see the *Model Patchwork Algorithm* section for further details). In order to augment and standardize the images, we used random crop, which created a standard sub-image with 128-by-128 pixels at a random location within the original fully-sampled image. This data augmentation step was performed every time the deep learning algorithm loaded a fully-sampled image to form a batch of sub-images for training. To stabilize training, as the random crop may land in a sub-image with too few blood vessels, we performed 10 random crops per iteration.

The other data augmentation operations include random rotation (up to 20 degrees), random lateral shift (up to 20 % of the image width), random vertical shift (up to 20 % of the image height), and random shear (up to shear factor of 0.2) [39]. Each of the augmentation methods used a constant fill value of 0. About 10% of the data was also augmented with additive Gaussian noise with zero mean and a standard deviation of 0.1 (subsequently being renormalized). All of the random augmentation techniques used a random seed of 7 for reproducible results. The validation dataset used a crop of 128-by-128 pixels extracted from the center of the image. Other than 10 random crops being used on the validation dataset to stabilize the performance of our saving metric (see *Network Training*), none of the data augmentation techniques were used on the validation and testing datasets.

### C. Downsampling Procedure

We artificially downsampled our fully-sampled PAM images (Fig. 3a) in order to mimic the undersampling performed in practice. For example, if the artificial scanning step size in the x-direction is five times as large as the fully-sampled step size, we downsample the x-axis by a ratio of 5:1. This method can be used to synthetically recreate different downsampling ratios. The downsampling ratio follows the format of [Sx, Sy], where Sx and Sy are the downsampling ratios along the x-axis and y-axis, respectively. Note the x and y directions are with respect to the images. For example, if an image is undersampled by a factor of 5 in the x-direction and 7 in the y-direction, the downsampling ratio is [5, 7].

For the downsampled images, we can add back in these missing pixels according to the downsampling ratios. We tested two different approaches to add back in these missing pixels. The first approach was to use zero-fill, in which missing pixels were replaced with a constant value of zero. The second approach tested was to resize the downsampled images using bicubic interpolation. After testing both of these techniques on images downsampled at a ratio of [7,3], we found that the zero-fill method outperformed the bicubic resizing by approximately 19% in the peak signal-to-noise ratio (PSNR) and 18% in the structural similarity index (SSIM) [40]. As such, we chose to move forward with the zero-fill method for resizing the downsampled images in all of our experiments.

### D. Network Training

In our work, all of the networks were optimized using the Adam algorithm [41] with a mini-batch size of 16. The Adam hyperparameters were set as: initial learning rate = 0.001, $\beta_1$ = 0.9, $\beta_2$ = 0.999, and $\varepsilon = 1 \times 10^{-7}$ (the default parameters of Tensorflow). As an optimization technique, Adam is generally considered quite robust to one's choice of hyperparameter values, so we kept the balanced default values provided by Tensorflow [42]. The models were trained for 500 epochs (~24 hours), with 10 random crops per image and the other aforementioned augmentations for the training images. The performance gains after training for approximately 350 epochs (~14.5 hours) are moderate (~1-2%) in some cases, but not necessary to achieve an effective model if time-constrained. For the batch normalization, the momentum parameter was set to 0.99 and ε was set to 0.001. For each of the models trained, a saving metric was used to determine when to set model checkpoints. The saving metric combines SSIM and PSNR as follows:

$$Saving\ Metric = (1 - SSIM) + \frac{40 - PSNR}{275} \quad (5)$$

During the training, the PSNR of the images never exceeded approximately 35.0; this is why 40.0 was used in Eqn. (5) as an empirical PNSR limit. The division by 275 was to ensure the SSIM and PSNR were on the same scale and that SSIM was weighted more heavily in the saving metric than PSNR.

The networks were implemented using Python 3.7.6 in Keras with a Tensorflow backend. The workstation setup included an AMD Ryzen 7 1700x CPU, 32 GB RAM, and a NVIDIA GTX 1080Ti.

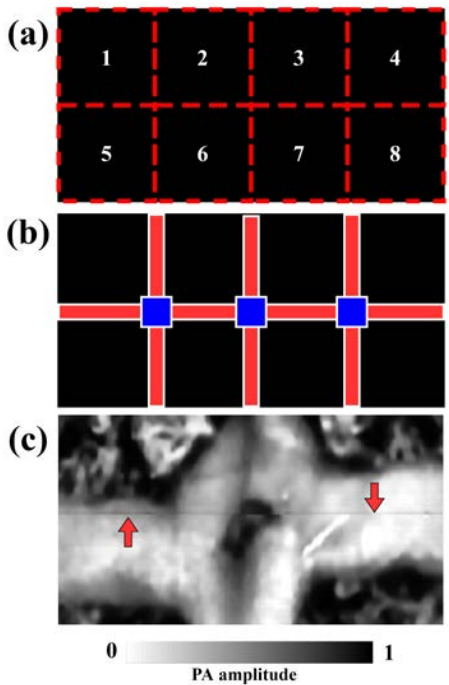

Fig. 3. Illustration of the patchwork algorithm for the first pass and subsequent cleaning passes. Due to edge aberrations that can occur in the first pass (numbered in (a) and colored black in (b)), we perform a second pass (shown in red in (b)) and a third pass (shown in blue in (b)) to remove the edge distortions visible in (c) as horizontal dark lines (see red arrows).

*E. Model Patchwork Algorithm*

In ideal conditions, our FD U-net model should take 128 × 128 pixels downsampled PAM images as the input and output 128 × 128 pixels images that approximate the fully-sampled ground truth. However, a full murine brain image, for example, might have at least 1500 × 1500 pixels, which cannot be directly fed into our model. To overcome the image size limitation of our model, we developed an additional algorithm to transform larger images into 128 × 128 pixel patches that could be processed by our model and stitched back into the original image size. Our patchwork algorithm pads both image axes with zeros such that the dimensions of the image are evenly divisible by 128. Next, the algorithm loops through the image to form non-overlapping patches of 128 × 128 pixels (Fig. 3 (a)).

In the first pass, each patch is fed into our model as **X** and the output **Y** (Fig. 1) is placed at the patch's original location within the full-size image. However, refinement is needed as each patch may contain artifacts on its edges (Fig. 3 (c)). These artifacts are removed by two subsequent cleaning passes. The second pass reprocesses the original downsampled image, but the patches are offset such that the patch center falls on the edges of the first pass patches (the red dotted lines in Fig. 3 (a)). The output of the second pass replaces the edge pixels of the first pass with a 20 pixel deep buffer zone (red regions in Fig. 3 (b)). The third pass then refines the horizontal-vertical intersection areas of the first pass patches, to eliminate the edges artifacts left from the first pass or introduced in the second pass (blue regions in Fig. 3 (b)). These three passes cover all possible edge aberration regions.

IV. RESULTS

*A. Model Architecture comparison*

In our preliminary model architecture comparison, we evaluated the performance of our modified FD U-net [11] with strided convolutions, a modified U-net [43] structure with identity activation [10], a modified U-net structure that adds residual connections during the compression stage (Res U-net) [44] and strided convolutions, as well as a modified Res U-net that also employs independent component layer (ICL) blocks (ResICL U-net) [45] in the compression phase. To compare these architectures, each was trained using solely MAE for 500 epochs with the Adam optimizer (initial learning rate = 0.001, $\beta_1$ = 0.9, $\beta_2$ = 0.999, and $\varepsilon$ = 1×10$^{-7}$). All of these model architectures were trained with the same training data downsampled at a ratio of 5:1 in the x-direction, meaning that only 20% of the original pixels were used. In other words, the effective imaging time could be reduced by 80%. After each model was trained for 500 epochs, we used a variety of loss metrics to compare their performance, as shown in Table I. FD U-net had the best performance in PSNR and MSE, and was second best in the SSIM. In addition, FD U-net showed the highest robustness, as the other models often required greater contrast rescaling due to the presence of single-pixel high-intensity model response aberrations. This lack of robustness is exemplified by U-net, which required very high levels of contrast rescaling to surpass a PSNR of 25.0 and an SSIM of 90.0. With the same contrast enhancement as our novel architectures and FD U-net, U-net was only able to achieve an SSIM of ~85.6 (primarily due to the presence of high-intensity model response aberrations). To keep things equal for all the models during the model architecture search, we clipped the pixel values below 0.05 percentile and above the 99.95 percentile, rescaling the remaining pixels to 0-1. These

TABLE I
STATISTICAL METRICS (MEAN ± SD) TO COMPARE MODEL ARCHITECTURES

|  | U-net | Res U-net | ResICL U-net | FD U-net | Bicubic Interpolation |
|---|---|---|---|---|---|
| PSNR | 24.06 ± 6.05 | 29.30 ± 1.56 | 29.36 ± 1.55 | **29.40 ± 1.56** | 25.79 ± 2.21 |
| SSIM | 0.8560 ± 0.0874 | **0.9176 ± 0.0213** | 0.9106 ± 0.0247 | 0.9157 ± 0.0215 | 0.8445 ± 0.0472 |
| MAE | 0.04850 ± 0.0406 | 0.01818 ± 0.00390 | **0.01769 ± 0.00361** | 0.01835 ± 0.00378 | 0.02671 ± 0.008622 |
| MSE | 0.009623 ± 0.012317 | 0.001261 ± 0.000538 | 0.001242 ± 0.000531 | **0.001232 ± 0.000524** | 0.002984 ± 0.001475 |

*PSNR*, peak signal-to-noise; *SSIM*, structural similarity index; *MAE*, mean absolute error; *MSE*, mean squared error

TABLE II
STATISTICAL METRICS (MEAN ± SD) TO COMPARE DOWNSAMPLING RATIOS

|  | Downsampling Ratio: [5, 1] | | Downsampling Ratio: [7, 3] | | Downsampling Ratio: [10, 5] | |
| --- | --- | --- | --- | --- | --- | --- |
|  | FD U-net | Bicubic Interpolation | FD U-net | Bicubic Interpolation | FD U-net | Bicubic Interpolation |
| PSNR | **30.4 ± 2.0** | 25.8 ± 2.2 | **26.9 ± 2.0** | 23.7 ± 2.2 | **23.6 ± 1.8** | 20.3 ± 2.0 |
| SSIM | **0.923 ± 0.019** | 0.845 ± 0.047 | **0.854 ± 0.038** | 0.780 ± 0.061 | **0.782 ± 0.046** | 0.651 ± 0.076 |
| MAE | **0.014 ± 0.004** | 0.027 ± 0.009 | **0.023 ± 0.006** | 0.035 ± 0.012 | **0.033 ± 0.009** | 0.055 ± 0.015 |
| MSE | **0.0010 ± 0.0006** | 0.0030 ± 0.0015 | **0.0023 ± 0.0010** | 0.0048 ± 0.0026 | **0.0047 ± 0.0018** | 0.010 ± 0.004 |

*PSNR*, peak signal-to-noise; *SSIM*, structural similarity index; *MAE*, mean absolute error; *MSE*, mean squared error thresholds were chosen to raise the performance of plain U-net to be comparable with bicubic interpolation. Upon testing various thresholds, the essential performance relationship of FD U-net, Res U-net, and ResICL U-net stayed relatively consistent. Due to its superior performance and robustness, as well as the fact that it is a more established model architecture, FD U-net was eventually selected in this work for the bulk of the experiments with different downsampling ratios. For more details on the various model architectures tested, please see Supplemental Figures 1-3.

### B. FD U-net performance at different downsampling ratios

Next, we investigated the performance of FD U-net at different downsampling ratios, against the performance of the linear upsampling method using bicubic interpolation. As the robust FD U-net architecture did not suffer as many contrast enhancement issues as the other models, we evaluated the model with minimal contrast enhancement at three representative downsampling ratios of [5, 1], [7, 3], and [10, 5]. The comparison of our model's performance to bicubic interpolation is shown in Fig. 4. As an example, the results of FD U-net and bicubic interpolation on a full mouse brain image with a downsampling ratio of [7, 3] are shown in Fig. 5. The model performance for all 29 test images is summarized by the statistics in Table II.

### C. Downsampling Ratio: [5, 1] – 20% effective pixels

We first examined a downsampling ratio of [5, 1], in which only 20% of the original pixels were used for the image upsampling. A sub-region of the fully-sampled image is shown in Fig. 5 (a2), while the downsampled image is depicted in Fig. 6 (a-I). The results from FD U-net and bicubic interpolation are shown in Fig. 6 (a-II) and (a-III) respectively. It is clear from the statistical results (Table II) that the deep learning model vastly outperformed bicubic interpolation. However, the differences in the image results can be subtle. A clear distinction between the two methods is the quality of "vesselness" (i.e., smoothness and roundness) in which FD U-net greatly outperformed bicubic interpolation. As visible in the profiles of vessels, especially in the small vessels, bicubic interpolation suffers from jagged and disjointed features that do not exist in the FD U-net reconstruction.

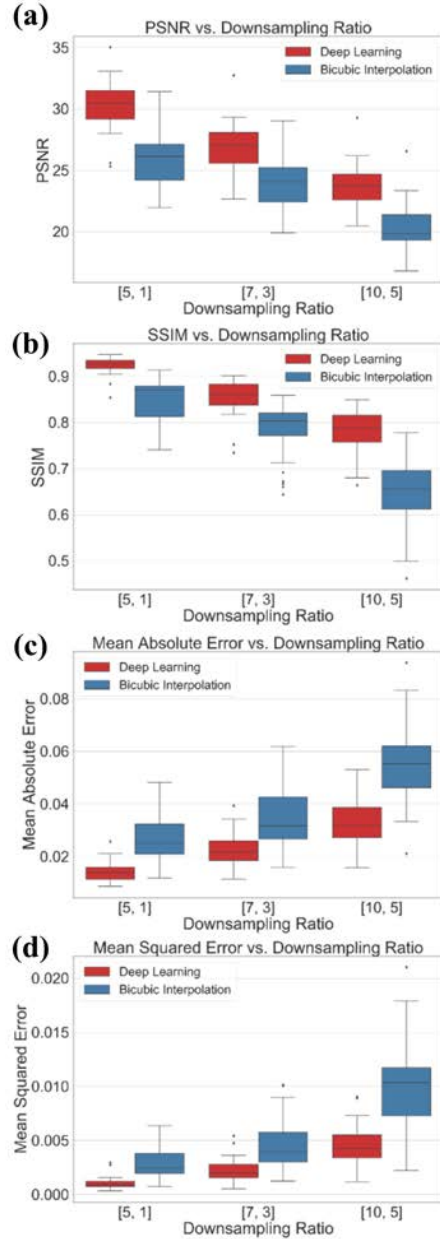

Fig. 4. Quantitative comparison of deep learning and bicubic interpolation upsampling performance at different downsampling ratios, in terms of (a) PSNR, (b) SSIM, (c) mean absolute error, and (d) mean squared error.

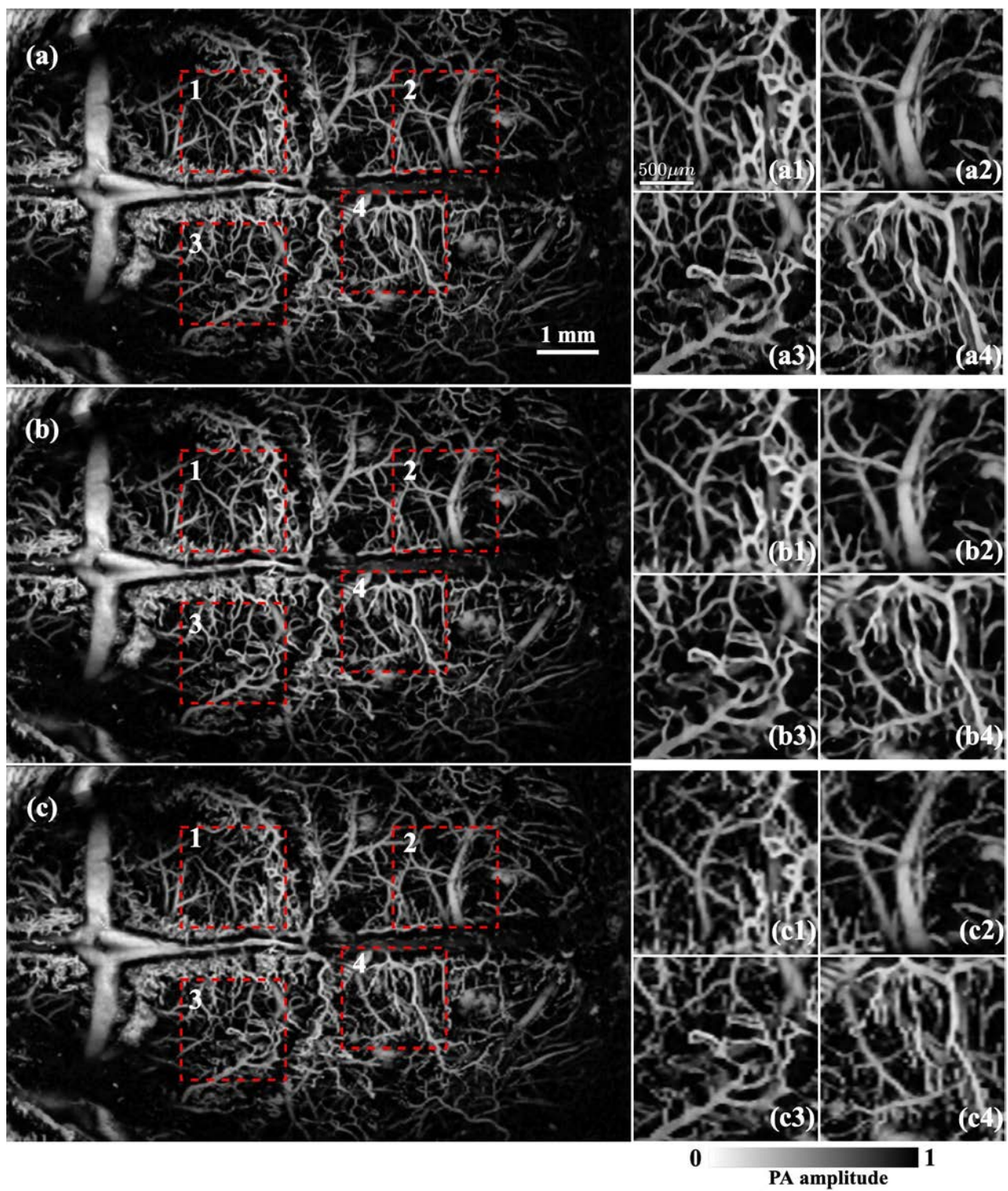

Fig. 5. The performance of FD U-Net compared to bicubic interpolation with a downsampling ratio of [7, 3]. (a) Fully-sampled whole-brain vascular image as the ground truth. Close-up images of the dashed box regions (1-4) are shown to the right as (a1)-(a4). (b) FD U-Net results from the downsampled data. (c) Bicubic interpolation results from the downsampled data.

*D. Downsampling Ratio: [7, 3] – 4.76% effective pixels*

The comparative metrics continue to show the superior performance of the FD U-net model in the [7, 3] downsampled data (Table II), with FD U-net's performance at this downsampling ratio actually exceeding the performance of bicubic interpolation at [5, 1]. In addition, the image quality differences between the two methods have become more significant at this sparsity. At the [7, 3] downsampling ratio, FD U-net's reconstruction in Fig. 5 (b) greatly outperforms bicubic interpolation in Fig. 5 (c) in terms of vesselness. Using less than 5% of the pixels, bicubic interpolation creates reconstructions with jagged and biologically improbable vessel profiles (Fig.

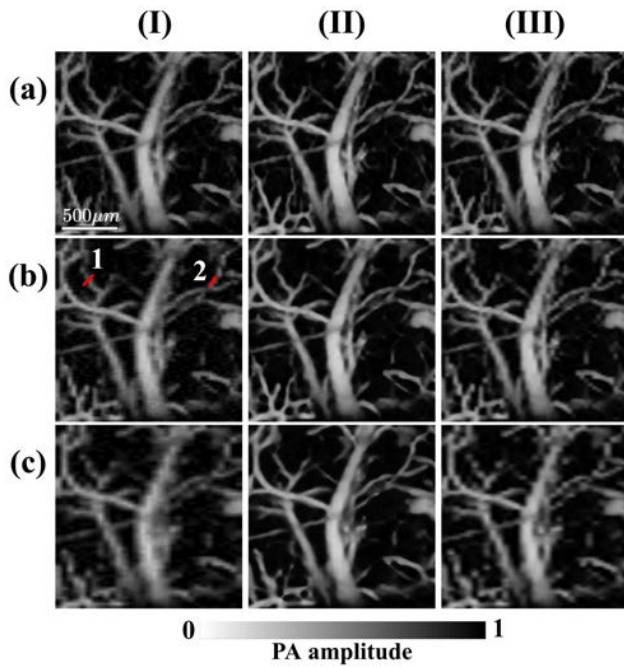

Fig. 6. Comparison of FD U-Net performance with different downsampling ratios of the full-sampled image in Fig. 5 (a2). At downsampling ratios of [5, 1], [7, 3], and [10, 5] (as rows (a), (b), (c) respectively), we have depicted (I) the downsampled images, (II) the FD U-Net reconstructions, and (III) the bicubic interpolation results.

7), while the deep learning model is able to reconstruct smoother vessels. This quality of vesselness is difficult to replicate with phantoms, which highlights our training strategy using fully-sampled *in vivo* data as the ground truth. The image quality of FD U-net at [7, 3] downsampling is still acceptable. Because the reconstruction used only 4.76% of the original pixels, we can expect a reduction in the imaging acquisition time by 95.24%

### E. Downsampling Ratio: [10, 5] – 2% effective pixels

The third downsampling ratio we tested was [10, 5], which constituted reconstruction from ~2% of the original pixels. We tested this downsampling ratio to explore the limit of the deep learning model. At [10, 5], as shown in Table II, the FD U-net still outperformed the bicubic interpolation in all the listed metrics. As shown in Fig. 6 (cII), the FD U-net model begins to blur vessels that are close together or overlapping. Although the performance of our model decreased at this very high level of downsampling, the bulk physiology of the microvasculature would be acceptable, given that the imaging speed can be potentially improved by 50 times.

## V. DISCUSSION

Our work builds on the recent innovations in deep learning, and applies these advances to the fast-growing PAM. By collecting and training on an increasing set of *in vivo* mouse brain microvascular data, our FD U-net model was able to learn how to reconstruct images at downsampling ratios of up to 50 times. Using between 2% and 20% of the original pixels, our FD U-net model can potentially accelerate the imaging speed of the traditional PAM systems with larger scanning step sizes. This approach circumvents the expensive hardware advances that are currently researched and, as our dataset grows, builds an avenue to even greater performance with continuous retraining and refinement. The three representative downsampling ratios enabled us to demonstrate the relative performance of the FD U-net model and bicubic interpolation as downsampling increased, and prove the superior capability of our learned model to retain the vesselness of the fully-sampled image. Our results show deterioration in performance as sparsity increases (Fig. 4), but this deterioration should always be put into context of the desired resolution and imaging speed of the experiment. The downsampling ratio can be tuned to fit specific applications as a balance between resolution and imaging speed. Our deep learning model expands the range of acceptable downsampling ratios into those that were previously forbidden in PAM.

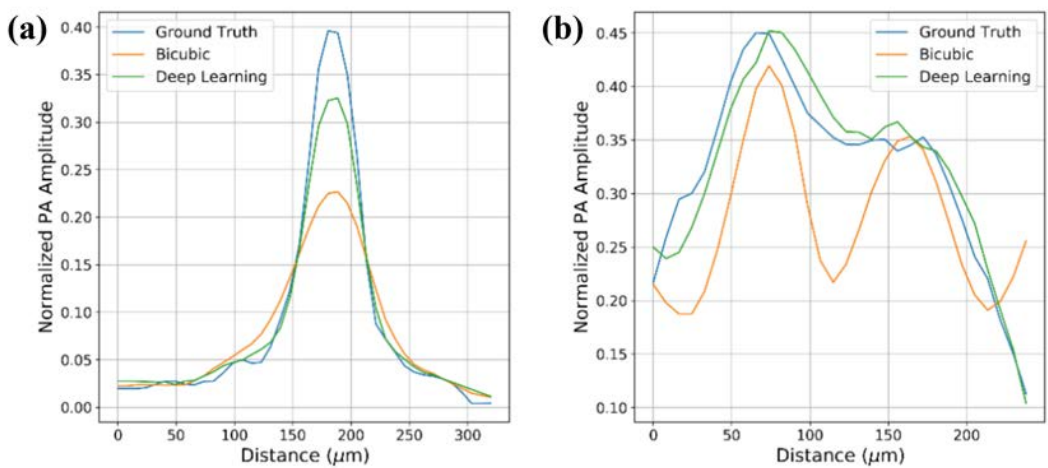

Fig. 7. Comparison of upsampling performance on two small vessels labeled by the dashed lines in Fig. 6 (b-I) as vessel 1 (a) and vessel 2 (b). The downsampling ratio used was [7, 3].

By training and testing on various downsampling ratios, we were also able to demonstrate the limitations of the current deep learning architectures and forge paths toward new innovations. Moving forward, a key improvement will be the refinement and application of new model architectures such as generative adversarial networks (GANs) [16, 42, 46], which may yield modest performance gains. In addition, we will develop tailored deep learning models for specific PAM system implementations. This may be done in conjunction with the technical advances in compressed sensing or fast-scanning mechanisms [2, 4, 5, 6, 7, 8]. Our method is also different from the traditional super-resolution imaging approaches, as we do not exceed the baseline resolution achievable by our system, but rather improve image quality with pixel-wise subsampling [47].

One of the most prominent challenges in implementing deep learning models in PAM has been the need to acquire a large amount of fully-sampled *in vivo* PAM data. Our lab hopes to be the first of many to take steps toward openly sharing PAM data for the benefit of the community [48]. At the end of the paper, researchers can find a link to the entire dataset used in our work as well as our source code. We aim to create and share a large database of PAM vascular images to which many researchers can contribute and use for their own machine learning applications. This database will grow continuously as our lab generates new data, allowing for continuous retraining and refinement. In addition, because of the high quality of PAM vessel data, it should be possible for researchers in other imaging fields such as two-photon microscopy and optical coherence tomography to train models using the PAM database.

## VI. CONCLUSION

Here, we have demonstrated a novel application of deep learning principles in order to address the trade-off of imaging speed and spatial resolution in undersampled PAM. We tested different model architectures and found that FD U-net has the best performance (Table I). Our modified FD U-net model architecture outperformed bicubic interpolation (Table II) at all of the representative downsampling ratios tested. By making our mouse brain microvasculature dataset and model source code freely available to the research community, we hope to maximize the impact of our PAM deep learning dataset.

## DATA AND CODE AVAILABILITY

All of the mouse brain microvasculature datasets used for this study were generated in our laboratory and are downloadable at the Open Science Framework (OSF). The main code used to produce the results in this paper is available on https://github.com/axd465/PAM_Deep_Learning_Undersampling_Publication.

## ACKNOWLEDGEMENTS

The authors thank Dr. Caroline Connor for editing the manuscript.

# SUPPLEMENTAL FIGURES:

## *Diagram 1 – U-net Architecture*

The variables $L_1, L_2, L_3, \ldots L_i$ refer to the level of compression depth (i) within the model and the image size (N × N) at that compression depth

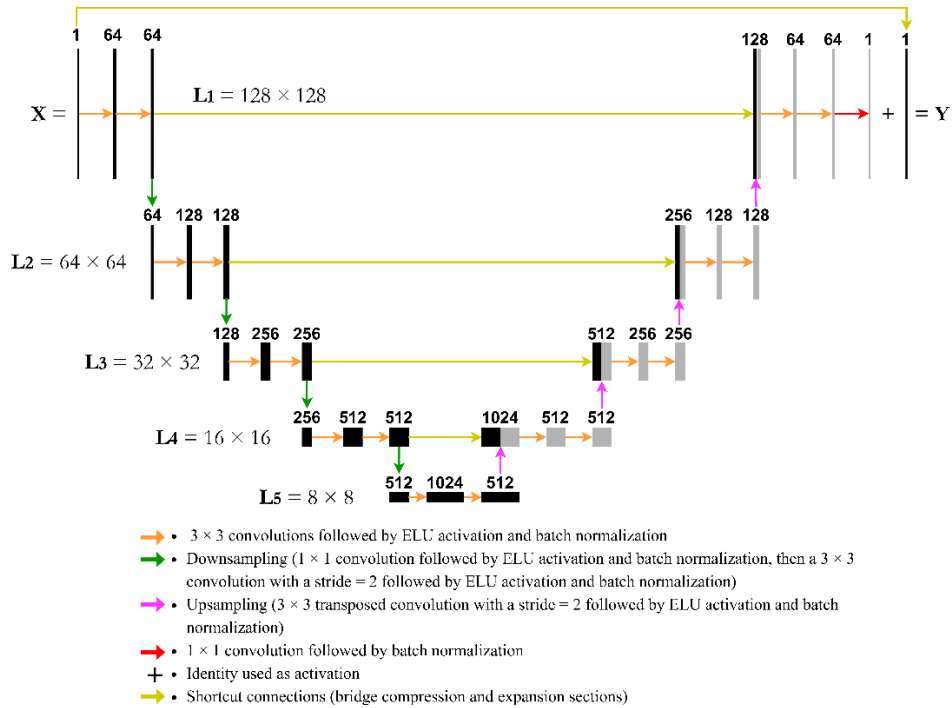

## *Diagram 2 – Res U-net Architecture*

The variables $L_1, L_2, L_3, \ldots L_i$ refer to the level of compression depth (i) within the model and the image size (N × N) at that compression depth

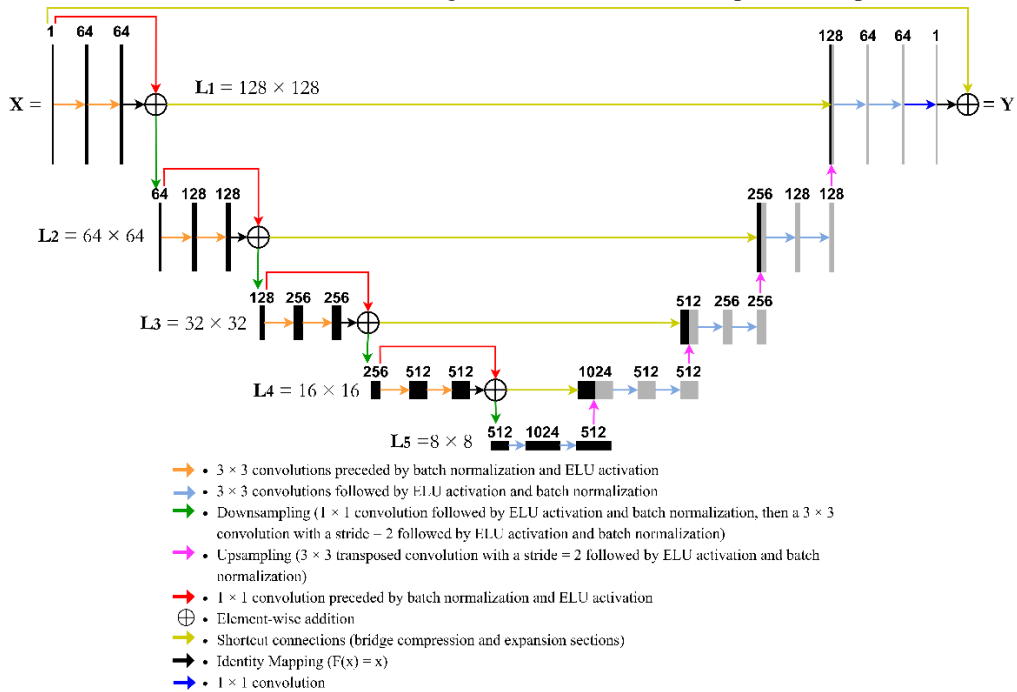

## Diagram 3 – ResICL U-net Architecture

The variables $L_1, L_2, L_3, \ldots L_i$ refer to the level of compression depth (i) within the model and the image size ($N \times N$) at that compression depth

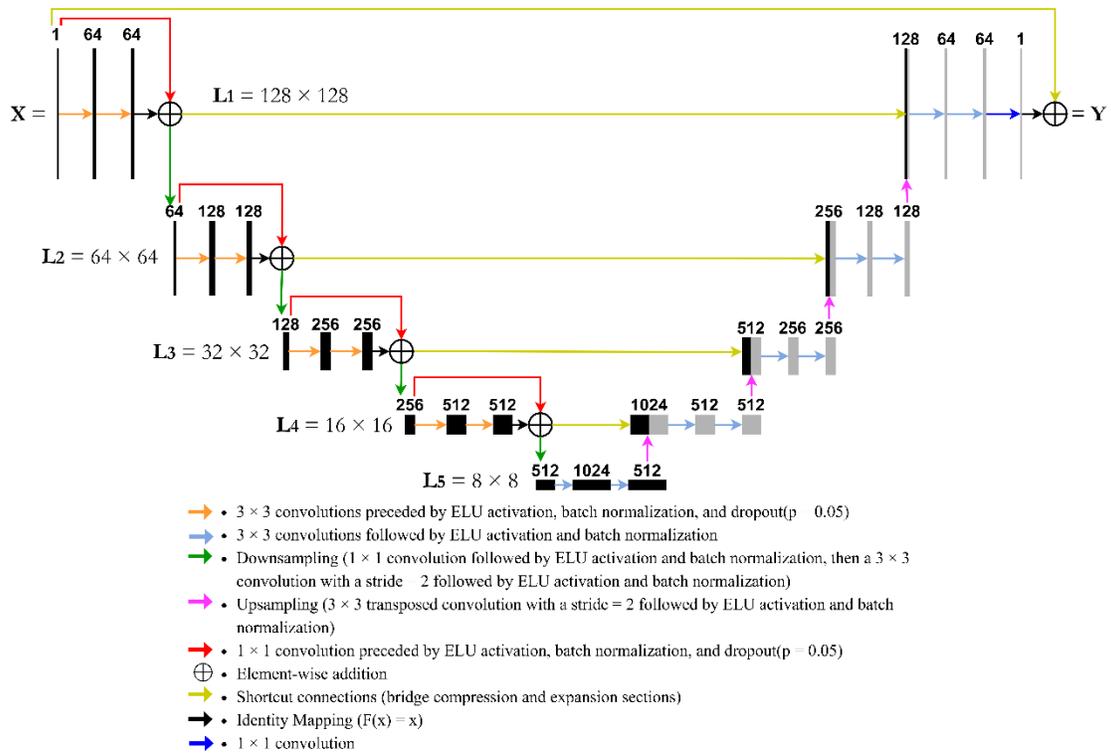